# Interference-governed electromagnetic-thermal coupling and heat transport in pulse EUV-irradiated multilayer nanofilms

Hongyu He[1,2,#], Li Ma[1,#], Zhiyi Xie[1], Yufan Liu[1], Chao Wu[3], Qiye Zheng[4], Yi Tao[5,*], Yunfei Chen[5], Chenhan Liu[1,2,*]

1. *Advanced Thermal Management Technology and Functional Materials Laboratory, School of Energy and Mechanical Engineering, Nanjing Normal University, Nanjing, 210023, P. R. China*
2. *Ministry of Education Key Laboratory of NSLSCS, Nanjing Normal University, Nanjing, 210023, P. R. China*
3. *School of Electrical and Automation Engineering, Nanjing Normal University, Nanjing, 210023, P. R. China*
4. *Department of Mechanical and Aerospace Engineering, The Hong Kong University of Science and Technology, Clear Water Bay, Kowloon, Hong Kong*
5. *Jiangsu Key Laboratory for Design and Manufacture of Micro-Nano Biomedical Instruments, School of Mechanical Engineering, Southeast University, Nanjing, 211100, P. R. China*

**# These authors contribute equally to this work**

***Corresponding authors:** chenhanliu@njnu.edu.cn; yitao@seu.edu.cn

## Abstract

Mo/Si multilayer mirrors are central to extreme ultraviolet lithography, where nanoscale optical interference and heat accumulation together constrain reflectivity and operational stability. Here we develop an analytical electromagnetic–thermal coupling model that directly links transfer-matrix-based interference-controlled energy deposition with transient heat conduction in EUV-irradiated multilayers. The model reveals a fundamental trade-off whereby increasing the multilayer period number enhances reflectivity but simultaneously elevates temperature by impeding heat dissipation. Interference-driven volumetric absorption further gives rise to pronounced axial temperature gradients and a post-pulse downward migration of the heat-flux maximum, a delayed-heating effect inaccessible to conventional surface-flux-based models. Systematic analysis establishes scaling laws connecting interfacial thermal resistance, beam size, and incident energy density to thermal confinement and temperature rise. By incorporating interfacial compaction kinetics, the model enables a quantitative assessment of mirror lifetime. This work offers a theoretical tool for thermal–optical co-design of multilayer nanostructures including EUV mirrors under pulsed irradiation across a wide spectral range.

## Introduction

Over the past several decades, advances in lithography technologies[1–3] have continuously driven the miniaturization of transistors in microchips. To sustain Moore's law as device scaling progresses toward increasingly smaller technology nodes[4–6], lithography has now entered the era of extreme ultraviolet (EUV, 13.5 nm), where the substantially reduced wavelength enables a marked improvement in patterning resolution[7,8].

At the core of EUV lithography are Mo/Si nanoscale multilayer mirrors[9,10], composed of tens of alternating Mo and Si layers with individual thicknesses of only a few nanometers[11]. The precisely engineered nanostructure provides high reflectivity at 13.5 nm (Supplementary Fig. S1) and is thus indispensable for high-resolution pattern transfer[2]. Despite the critical role, the operational stability of Mo/Si multilayer mirrors is severely challenged under high-intensity EUV irradiation. Achieving the required optical performance necessitates nanometer-level control over individual layer thicknesses together with millimeter-scale thickness uniformity across the mirror surface; however, continuous energy deposition during irradiation leads to heat accumulation within the multilayer stack[12], inducing interfacial thermal compaction and structural degradation[13–15], which in turn results in a pronounced deterioration of optical performance[16,17]. Therefore, a quantitative understanding of the temperature evolution within Mo/Si multilayers is essential for overcoming this thermal bottleneck and improving mirror reliability under realistic operating conditions.

In nanoscale multilayer structures, electromagnetic absorption and the ensuing heat diffusion exhibit pronounced spatial and temporal heterogeneity, arising from nanoscale interference effects, material contrasts, and ultrathin layer geometries[18]. These complexities are inadequately captured by conventional thermal modeling approaches[19,20], which often rely on simplified absorption profiles or homogenized material descriptions. This limitation highlights the urgent need for a robust theoretical framework capable of resolving spatiotemporal temperature fields within EUV multilayers. Such framework is crucial for establishing a rigorous physical basis to assess thermal stability under high-flux EUV exposure and for guiding the rational design and optimization of next-generation multilayer mirror architectures.

Existing theoretical approaches, including the transfer matrix method[21,22] and multilayer heat transfer models (MHTM)[19,20], treat electromagnetic propagation and thermal diffusion as decoupled processes. This separation neglects the intrinsic coupling between electromagnetic loss distribution and subsequent heat transport, thereby limiting their ability to accurately predict transient thermal responses and temperature fields under EUV irradiation. Finite-element methods, while widely employed for multiphysics simulations across diverse physical systems[23,24], face severe challenges when applied to EUV multilayer mirrors. First, resolving nanometer-scale layer thicknesses simultaneously with millimeter-scale in-plane dimensions within a single finite-element mesh necessitates an extraordinarily large number of elements, leading to prohibitive memory requirements and computational cost[25,26]. Second, the characteristic length scales governing electromagnetic absorption and thermal transport are inherently mismatched. While EUV electromagnetic fields exhibit strong nanoscale modulation due to interference and standing-wave effects within the multilayer stack, heat diffusion evolves over much longer spatial and temporal scales, making it difficult for a unified finite-element mesh to simultaneously resolve both phenomena. Due to these limitations, FEM also struggles to provide reliable and efficient predictions of transient temperature fields in EUV multilayer mirrors. Overall, a unified, efficient, and cross-scale framework that self-consistently couples electromagnetic absorption with thermal evolution remains absent.

In this work, we develop an analytical cross-scale electromagnetic–thermal coupling model (ETCM) that self-consistently captures the interplay between interference-controlled electromagnetic absorption and transient heat diffusion in nanoscale multilayers. The ETCM enables efficient and accurate prediction of photothermal responses under pulsed EUV irradiation, while providing quantitative insight into how external irradiation conditions and intrinsic interfacial properties jointly govern thermal stability and structural evolution. Although established and validated here for Mo/Si multilayers under 13.5-nm EUV irradiation, the underlying coupling strategy—linking transfer-matrix-derived volumetric energy deposition with multilayer heat conduction—is fundamentally general. It can be directly extended to other multilayer optical stacks operating across a wide spectral range, from EUV to X-ray, and under diverse pulsed irradiation conditions. This framework therefore provides a rigorous theoretical basis for elucidating irradiation-induced degradation mechanisms in multilayer nanofilms and offers a practical route toward the thermal–optical co-design of multilayer optic systems.

## Results and Discussion

### Electromagnetic–thermal coupling model

In this work, we build the analytical model of interference-driven electromagnetic–thermal coupling based on a typical multilayer EUV mirror, schematically depicted in Fig. 1(a). The structure consists of a periodic Mo/amorphous-Si (a-Si) multilayer stack deposited on a crystalline-Si (c-Si) substrate, with a thin Ru capping layer on top. Notably, the proposed coupling framework is readily extendable to other multilayer-based optical systems. Conventional MHTM often treats the incident laser irradiation as a constant surface heat flux, as suggested in the inset of Fig. 1(a). However, in such periodic optical structures, the incident electromagnetic wave undergoes significant interference, which governs the spatial distribution of electromagnetic absorption and the resulting non-uniform thermal response. Therefore, an accurate model must explicitly account for the wave propagation and interference within the multilayer stack. To this end, the transfer matrix method is employed to describe the electromagnetic distributions, in which the electromagnetic field is decomposed into downward- and upward-propagating waves, as follows:

$$\begin{bmatrix} E_{j,t}^{+}(r,t) \\ E_{j,t}^{-}(r,t) \end{bmatrix} = G_{j(j-1)} P_{j-1} G_{(j-1)(j-2)} \ldots \ldots G_{21} P_1 G_{10} \begin{bmatrix} E_{0,b}^{+}(r,t) \\ E_{0,b}^{-}(r,t) \end{bmatrix}$$

$$= \begin{bmatrix} A_j & B_j \\ C_j & D_j \end{bmatrix} \begin{bmatrix} E_{0,b}^{+}(r,t) \\ E_{0,b}^{-}(r,t) \end{bmatrix}, \quad (1)$$

where $E_{j,t/b}^{+}$and $E_{j,t/b}^{-}$ represent the downward- (superscript –)and upward- (subscript +) propagating field amplitudes at the top (subscript $t$) or bottom (subscript $b$) boundaries of $j$-th layer (Fig. 1b), respectively, $G_{j(j-1)}$ denotes the transfer matrix at the interface between the ($j$-1)-th and $j$-th layers, and $P_j$ represents the propagation matrix within the $j$-th layer[27], with the explicit form given as follows:

$$P_j = \begin{bmatrix} e^{ik_j d_j} e^{-d_j/\delta_j} & 0 \\ 0 & e^{-ik_j d_j} e^{d_j/\delta_j} \end{bmatrix}, G_{j(j-1)} = \begin{bmatrix} \frac{1+\Delta_{j(j-1)}}{2} & \frac{1-\Delta_{j(j-1)}}{2} \\ \frac{1-\Delta_{j(j-1)}}{2} & \frac{1+\Delta_{j(j-1)}}{2} \end{bmatrix}. \quad (2)$$

Here, $\Delta_{j(j-1)} = u_{\mathrm{j-1}}/u_{\mathrm{j}}$ denotes the refractive index ratio at the interface, where $u_j$ is the refractive index and $d_j$ the thickness of the $j$-th layer, $\delta_j$ is the penetration depth of the electromagnetic wave in the $j$-th layer, $k_j$ is the wavevector of the electromagnetic wave in the $j$-th layer, related to the incident wavelength $\lambda$ and refractive index $u_j$ by $k_j = 2\pi u_j/\lambda$. When the wave reaches the bottom *c*-Si substrate (i.e., $n$-th layer), the reflected field at the bottom interface can be neglected because the penetration depth is far smaller than the thickness of bottom layer, yielding the boundary condition $E_{n,t}^{-} = 0$. Imposing this condition on Eq. (1) yields the relation $E_{0,b}^{-}/E_{0,b}^{+} = -C_n/D_n$. Thus, once $E_{0,b}^{+}$ is known from the incident electromagnetic intensity $Q$, $E_{0,b}^{-}$ can be obtained. With both $E_{0,b}^{+}$ and $E_{0,b}^{-}$ known, Eq. (1) directly gives the electric-field amplitude at each interface. Consequently, the downward- and upward-propagating field components within the $j$-th layer can be expressed as $E_j^{+}(z) = e^{ik_j z} e^{-z/(2\delta_j)} E_{j,t}^{+}$ and $E_j^{-}(z) = e^{-ik_j z} e^{z/(2\delta_j)} E_{j,t}^{-}$, respectively, where $z$ represents the position within the $j$-th layer.

Based on the solved electric-field distribution, the electromagnetic energy flux in the $j$-th layer can be expressed through the Poynting vector formulation[28] as follows:

$$\bar{S}_j(r,z,t) = \frac{1}{2}\sqrt{\frac{\varepsilon}{\mu}} \left| E_j^{+}(r,z,t) + E_j^{-}(r,z,t) \right|^2, \quad (3)$$

where $\mu$ and $\varepsilon$ denote the magnetic permeability and dielectric permittivity,

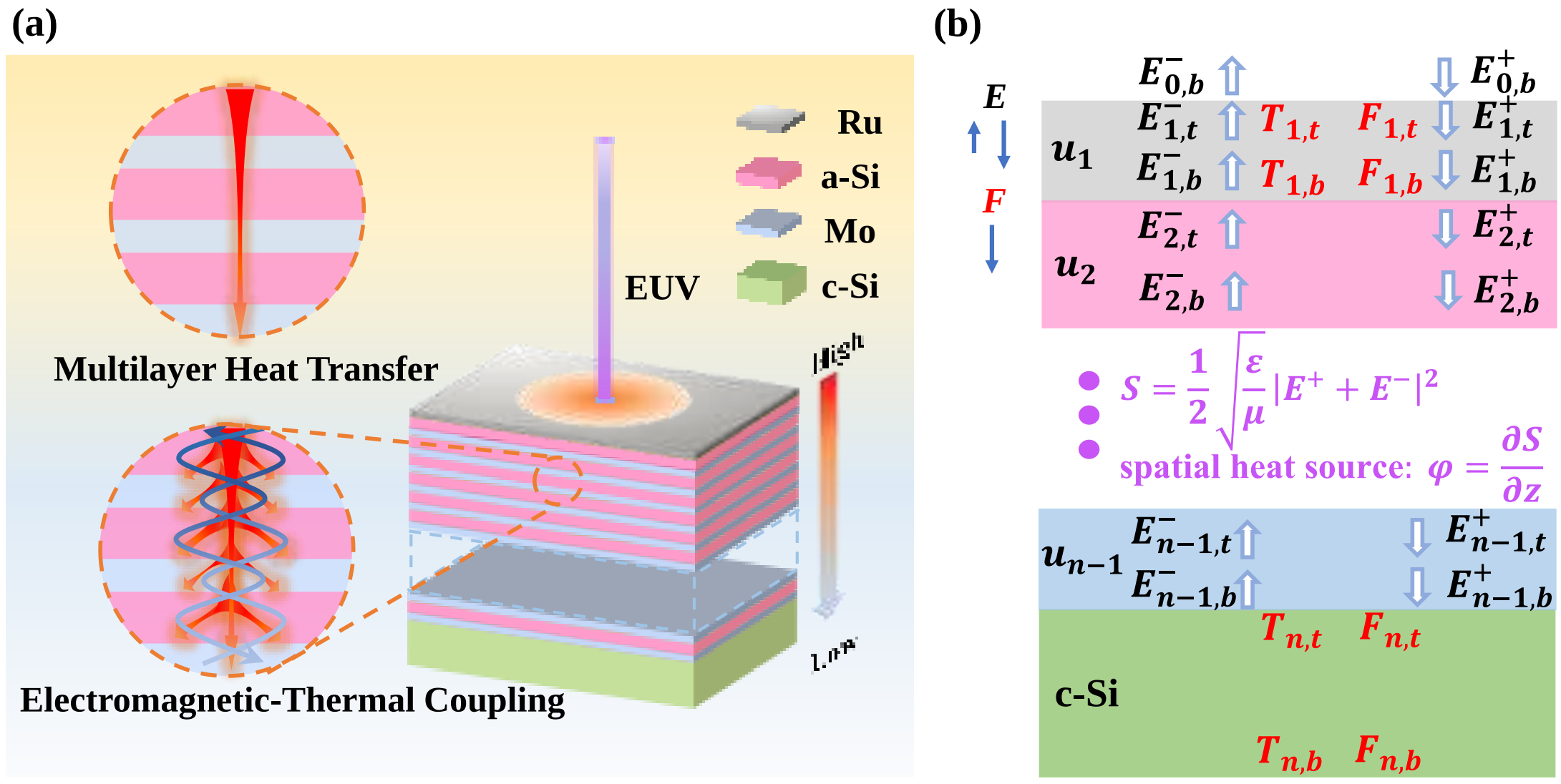

**Fig. 1 | Schematic of the electromagnetic-thermal coupling model and multilayer heat transport.** **a** Schematic illustration of the electromagnetic-thermal coupling framework and the multilayer heat transfer model, where the top inset enlarges the multilayer heat transport model, the bottom inset highlights the coupling model, and the blue curve represents the incident electromagnetic wave. **b** Distributions of the electromagnetic field, temperature, and heat flux within the multilayer stack.

respectively. As the electromagnetic wave propagates through the multilayer, a portion of its energy is absorbed and converted into a volumetric heat source within each layer. According to energy conservation, the volumetric heat source in each layer is proportional to the attenuation of the local electromagnetic flux[29], expressed as $\varphi_j(r,z,t) = \partial \bar{S}_j(r,z,t)/\partial z$. This volumetric heat source $\varphi_j(r,z,t)$ then acts as the source term in the transient heat conduction equation [30–33] :

$$C_{j,p}\frac{\partial\theta(r,z,t)}{\partial t} = \kappa_{j,z}\frac{\partial^2\theta(r,z,t)}{\partial z^2} + \frac{\kappa_{j,r}}{r}\frac{\partial}{\partial r}\left(r\frac{\partial\theta(r,z,t)}{\partial r}\right) + \varphi_j(r,z,t), \quad (4)$$

where $C_{j,p}$ is the volumetric heat capacity of the $j$-th layer, $\theta(r,z,t)$ is the temperature distribution, $\kappa_{j,z}$ and $\kappa_{j,r}$ are the axial and radial thermal conductivities of the $j$-th layer, respectively.

To obtain the transient temperature field under laser irradiation, the transient heat conduction equation (Eq. (4)) is transformed using a Fourier transform in time ($t \to \omega$) and a Hankel transform in the radial coordinate ($r \to l$). Applying the Fourier-Hankel-domain heat conduction equation and interfacial continuity conditions to each layer yields recursive relations for the temperature and heat flux as follows[19,20]:

$$\begin{bmatrix}\theta_{j,b}\\F_{j,b}\end{bmatrix} = M_j\begin{bmatrix}\theta_{j,t}\\F_{j,t}\end{bmatrix} + N_j = M_j O_{j(j-1)}\begin{bmatrix}\theta_{(j-1),b}\\F_{(j-1),b}\end{bmatrix} + N_j$$

$$= \underbrace{M_j O_{j(j-1)} \dots M_2 O_{21} M_1}_{\begin{pmatrix}U_j & V_j\\ Z_j & W_j\end{pmatrix}}\begin{bmatrix}\theta_{1,t}\\F_{1,t}\end{bmatrix} + \underbrace{M_j O_{j(j-1)} \dots M_2 O_{21} N_1 + \dots + N_j}_{\begin{bmatrix}X_j\\Y_j\end{bmatrix}}, \tag{5}$$

where $\theta_{j,t/b}$ and $F_{j,t/b}$ represent the temperature and heat flux at the top (subscript $t$) or bottom (subscript $b$) boundaries of $j$-th layer, $M_j$ denotes the thermal propagation matrix in the $j$-th layer, $N_j$ corresponds to the heat source matrix in the corresponding layer, and $O_{j(j-1)}$ represents the thermal interface condition between the $j$-th and ($j$-1)-th layers. These matrices are explicitly given by:

$$M_j = \begin{bmatrix}\cosh(p_j d_j) & -\frac{\sinh(p_j d_j)}{\kappa_{j,z} p_j}\\ -\kappa_{j,z} p_j \sinh(p_j d_j) & \cosh(p_j d_j)\end{bmatrix}, O_{j(j-1)} = \begin{bmatrix}1 & -\frac{1}{\sigma_{j(j-1)}}\\ 0 & 1\end{bmatrix}, \tag{6}$$

$$N_j = \begin{bmatrix}-\eta_j(l,0,\omega)\cosh(p_j d_j) - \frac{\partial \eta_j(l,0,\omega)}{\partial z}\frac{\sinh(p_j d_j)}{p_j} + \eta_j(l,d_j,\omega)\\ \kappa_{j,z} p_j \eta_j(l,0,\omega)\sinh(p_j d_j) + \kappa_{j,z}\cosh(p_j d_j)\frac{\partial \eta_j(l,0,\omega)}{\partial z} - \kappa_{j,z}\frac{\partial \eta_j(l,d_j,\omega)}{\partial z}\end{bmatrix}, \tag{7}$$

where $p_j = \sqrt{(\kappa_{j,r} l^2 + \mathrm{i} C_{j,p}\omega)/\kappa_{j,z}}$ [34], $\sigma_{j(j-1)}$ denotes the interfacial thermal conductance between the $j$-th and ($j$-1)-th layers, and $\eta_j$ represents the particular solution of Eq. (6) in the Fourier-Hankel domain. Since the substrate is assumed as semi-infinite, the corresponding bottom heat flux can be treated as zero, i.e., $F_{n,b} = 0$. Meanwhile, because all heat generation occurs internally via electromagnetic absorption, there is no independent heat flux incident on the top surface, resulting in the adiabatic surface boundary condition as $F_{1,t} = 0$. Applying these boundary conditions of $F_{n,b} = 0$ and $F_{1,t} = 0$ to Eq. (5) yields the top-surface temperature $\theta_{1,t} = -Y_n/Z_n$. These surface temperature $\theta_{1,t}$ and heat flux $F_{1,t}$ then serve as the starting point for a recursive calculation using Eq. (5), which yields the propagated temperature $\theta_{j,t}$ and heat flux $F_{j,t}$ at the top interface of each layer.

Using the solved temperature $\theta_{j,t}$ and heat flux $F_{j,t}$ as boundary conditions, the heat conduction equation in the Fourier-Hankel-domain can be solved to obtain the temperature distribution at any position *z* within the layer:

$$\theta_j(l,z,\omega) = \left(\theta_{j,t} - \eta_j(l,0,\omega)\right)\cosh(p_j z) + \left(-\frac{F_{j,t}}{\kappa_{j,z} p_j} - \frac{\partial \eta_j(l,0,\omega)}{\partial z}\frac{1}{p_j}\right)\sinh(p_j z) + \eta_j(l,z,\omega) \tag{8}$$

Then, the temperature distribution in the original spatiotemporal domain is recovered by applying the inverse Fourier transform ($\omega \to t$) and the inverse Hankel transform ($l \to r$) to Eq. (8), yielding:

$$\theta_j(r,z,t) = \frac{1}{2\pi}\int_{-\infty}^{\infty}\int_0^{\infty} \theta_j(l,z,\omega) J_0(lr) l e^{i\omega t} dl\, d\omega. \tag{9}$$

Using this equation, the spatiotemporal temperature distribution in the nanoscale Mo/Si multilayer structure under EUV irradiation can be analytically calculated.

**Model validation and benchmarking**

To assess the reliability of the proposed ETCM, we simulate pulsed EUV laser irradiation at 13.5 nm, corresponding to the peak reflectance of the Mo/Si multilayer stack. As a representative case, an 8 ns square-wave pulse with a repetition period of 200 ns is considered, with the corresponding temporal profile and optical response shown in Supplementary Figs. S1-S2. The laser energy density $Q$ is set to 0.5 $mJ/cm^2$. The multilayer optical structure and material parameters employed in the simulations are summarized in Supplementary Table S1, with the procedures for parameter evaluation detailed in Supplementary Note 2. A beam radius of 5 μm is adopted to enable direct benchmarking against the conventional MHTM and COMSOL simulations, for which the computational cost becomes prohibitive at larger spatial scales.

First, the temporal evolution and spatial profile of temperature predictions from the proposed ETCM are compared with those from a conventional MHTM in Fig. 2(a) and Fig.2 (b). While both models reproduce the overall temperature evolution, pronounced discrepancies are observed in the absolute temperature values. These deviations originate from the MHTM's neglect of electromagnetic wave propagation and interference within individual layers[20,35]. By treating the incident energy as a simple surface flux, the MHTM fails to capture the internal energy redistribution, leading to a systematic overestimation of temperature field. Notably, MHTM is a special case of ETCM proposed here. When the volumetric heat source $\varphi_j(r,z,t)$ in Eq. (4) is reduced to an equivalent surface heat source, the ETCM formally collapses to the MHTM, as rigorously derived in Supplementary Note 1. This well-defined reduction establishes a clear theoretical connection between the two models and provides an internal consistency check, thereby benchmarking and validating the ETCM developed in this work.

To further validate the ETCM, its predictions are compared against multiphysics simulations performed in COMSOL, as illustrated in Fig. 2(c-e). As shown in Fig. 2c, the electric-field distributions predicted by the ETCM and COMSOL are identical, exhibiting excellent agreement in both amplitude and period. This match confirms that ETCM accurately capture the optical interference effects within the multilayer structure[36–38]. Consistent agreement is also observed in the coupled thermal response, including the transient temperature evolution in Fig. 2(d) and the spatial temperature profile at the end of the pulse (8ns) in Fig. 2(e) (the radial distribution is provided in Supplementary Fig. S3, which also shows good agreement).

The computational efficiency of the proposed ETCM against the COMSOL is evaluated in Fig. 2(f). As the lateral size of the COMSOL simulation domain increases from 25 μm to 100 μm, the computation time rises substantially, consistent with the growth of finite-element degrees of freedom. In contrast, the ETCM employs a fully analytical formulation that eliminates spatial meshing and iterative numerical solvers, enabling direct evaluation of the temperature field for infinite lateral dimensions. As a result, the ETCM achieves a speedup of approximately 48× compared with COMSOL simulations at a 100 μm lateral size. This high computational efficiency facilitates rapid exploration of large parameter spaces and systematic thermal stability assessments, which would be impractical using the finite-element approaches.

As highlighted in the inset of Fig. 2(e), the temperature gradients in the Ru and Mo layers are relatively small, reflecting rapid heat diffusion due to their high thermal conductivities, whereas more pronounced variations in the amorphous Si layers suggest localized heat accumulation[39]. Meanwhile, due to the interfacial thermal resistance, the

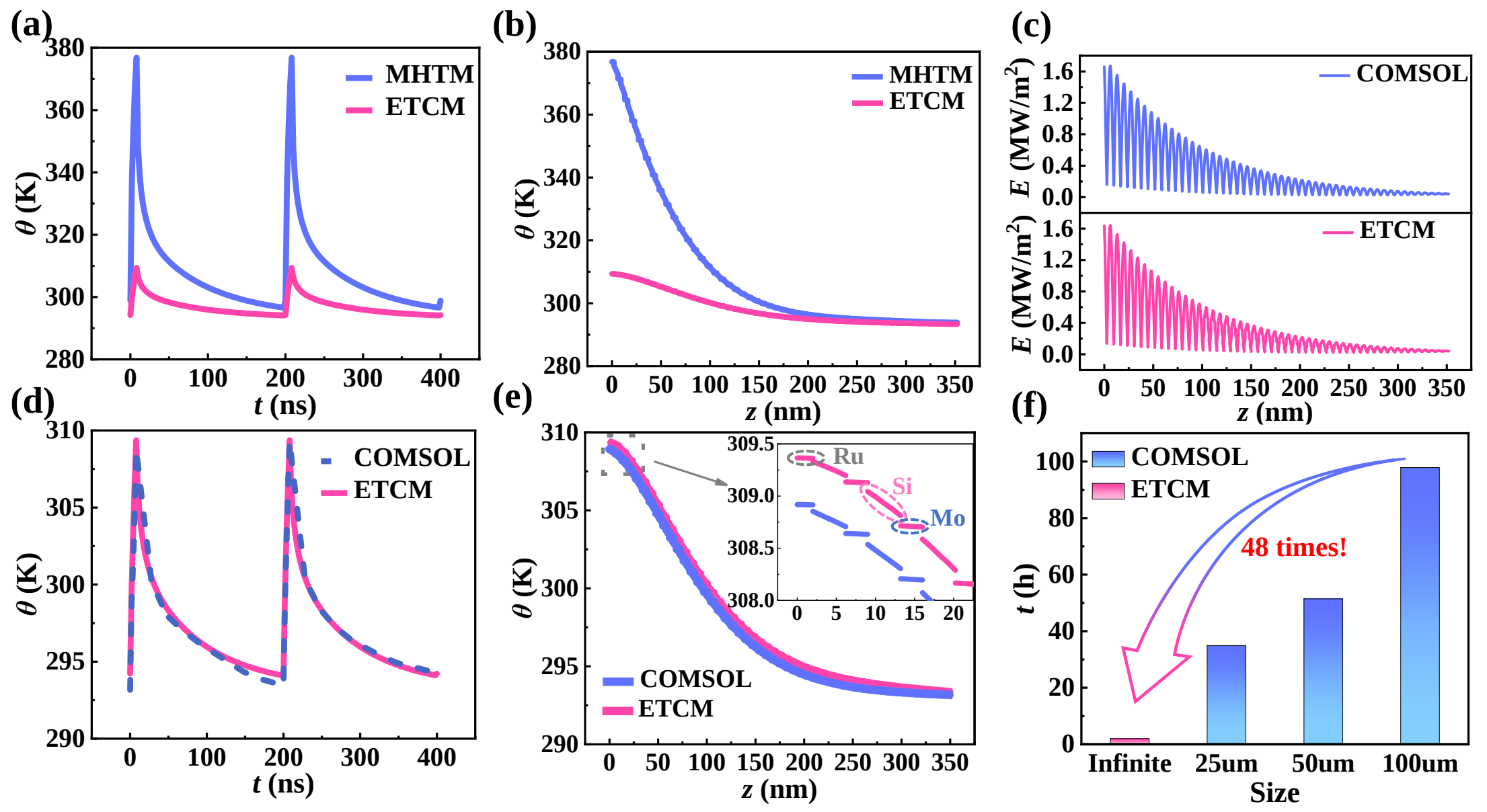

**Figure 2 | Validation and benchmarking of the electromagnetic–thermal coupling model.**

**a** Temporal temperature evolution at the film surface center predicted by ETCM and MHTM models. **b** Spatial temperature profiles of two models along the film thickness at the central point. Trends agree, but differences in magnitude highlight the ETCM model's improved accuracy. **c** Center electric field along the film thickness, showing excellent agreement between the present model and COMSOL simulations. **d** Temporal temperature evolutions of COMSOL simulations and ETCM model under pulsed (square-wave) irradiation. **e** Spatial temperature profiles of COMSOL simulations and ETCM model along the thickness. **f** Computational time for COMSOL simulations with different model sizes and ETCM model, showing that the ETCM model is significantly faster than COMSOL simulations.

temperature jumps at the interfaces, hindering the heat diffusion.

Overall, comparisons with both the conventional MHTM and full multiphysics simulations in COMSOL collectively validate the accuracy of the proposed ETCM. Relative to the MHTM, the ETCM explicitly incorporates the photothermal coupling process by accounting for volumetric heat generation and electromagnetic interference effects, thereby enabling a more faithful description of the spatiotemporal temperature distribution within nanoscale multilayer films. In contrast to COMSOL-based finite-element simulations, the ETCM achieves substantially higher computational efficiency while retaining quantitative accuracy. More importantly, as discussed above, due to mismatch between the characteristic length scales governing electromagnetic absorption and thermal transport, the finite-element model often fails to yield a convergent solution. Together, these advantages position the ETCM as a robust and accurate framework for predictive thermal analysis of nanoscale multilayer systems under EUV irradiation.

**From interference-driven absorption to thermal design**

By providing this reliable framework, the validated ETCM enables a systematic investigation into the fundamental mechanisms governing the electromagnetic energy deposition and thermal response of Mo/Si multilayers under EUV pulsed irradiation. In the following studies, a more practical beam radius of 30 μm is adopted unless otherwise specified. As illustrated in Fig. 3(a), the electromagnetic-field distribution exhibits clear interference fringes due to multiple reflections at the interfaces, giving rise to an intensity pattern that alternates with depth while undergoing gradual attenuation. Furthermore, the resultant absorption energy distribution in Fig. 3(b), derived from this interference pattern, decreases overall with depth and exhibits noticeable oscillations across the multilayer stack. These features arise from the interplay of interference and absorption, in which the interference modulates the local field strength and thus leads to the oscillations, while material absorption accounts for the monotonic energy decay with depth. This combined effect produces a non-uniform absorption profile that directly determines the spatial variation of the electromagnetic field and the resultant temperature.

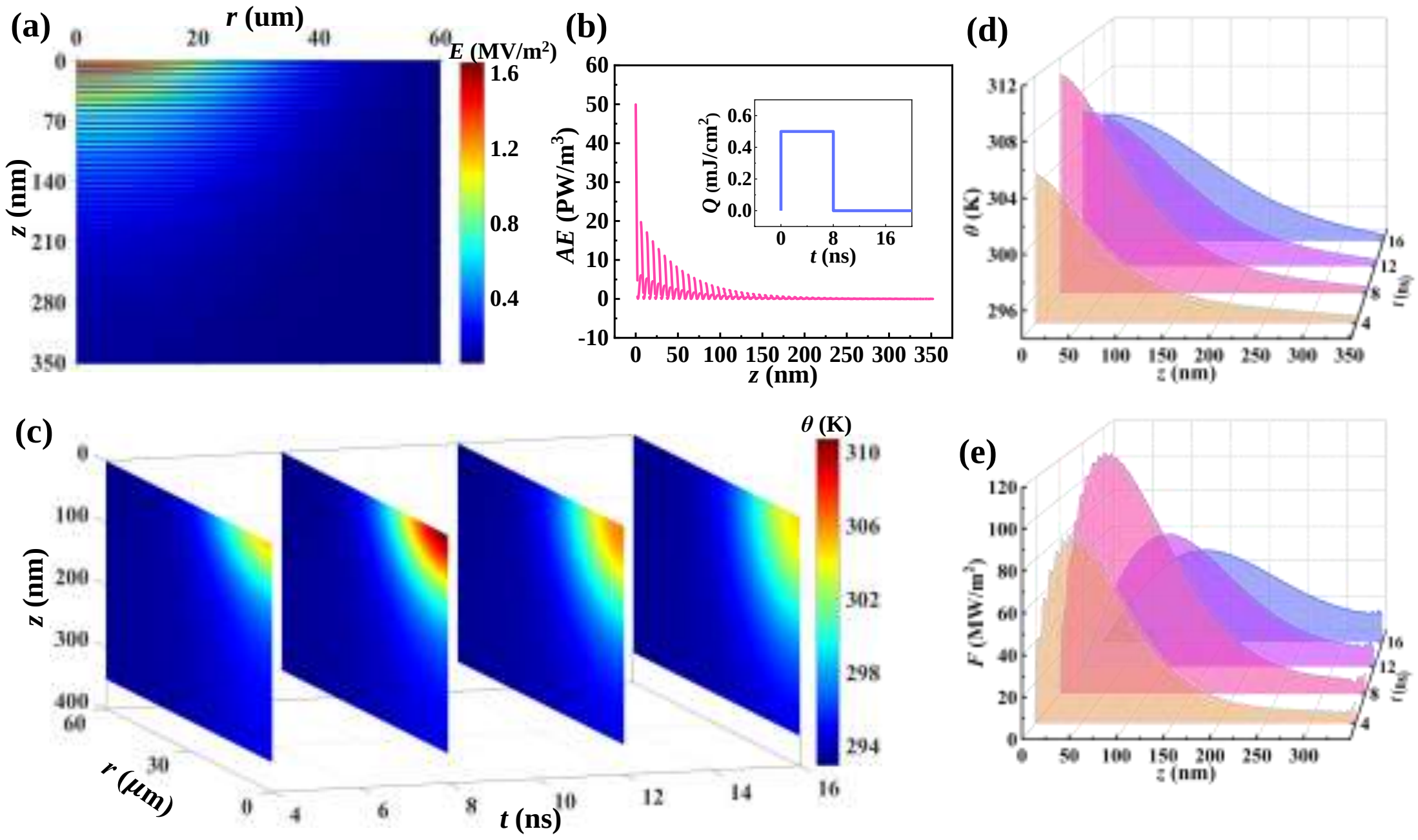

**Fig. 3 | Calculation results of ETCM model.**

**a** Electric field distribution, exhibiting pronounced interference effects. **b** Absorption profile along the axial ($z$) direction, with the inset showing the temporal waveform of the excitation pulse within the first 16 ns. **c** Spatiotemporal evolution of the temperature field at 4 ns, 8 ns, 12 ns, and 16 ns, illustrating transient thermal diffusion. **d** Axial temperature profiles over time. **e** Temporal evolution of axial heat flux, showing a diffusion trend with the heat flux peak gradually moving deeper into the multilayer structure and decreasing in magnitude.

This non-uniform absorption profile (Fig. 3(b)) acts as the volumetric heat source, directly giving rise to the spatiotemporal temperature evolution as shown in Fig. 3(c). The initial thermal response features a sharp temperature rise at the surface during the 0-8ns pulse, followed by slow radial thermal diffusion once the pulse ends, which is further illustrated by the time-resolved radial temperature profiles at the surface layer in Supplementary Fig. S4. In contrast, the axial temperature evolution in Fig. 3(d) progresses much more rapidly and in a more complex manner. During the 8-ns pulse, the temperature rises simultaneously throughout the multilayer depth. Once the pulse ends, the surface begins to cool, yet temperatures in deeper layers continue to increase, indicating a delayed downward propagation of the thermal front. This diffusion process

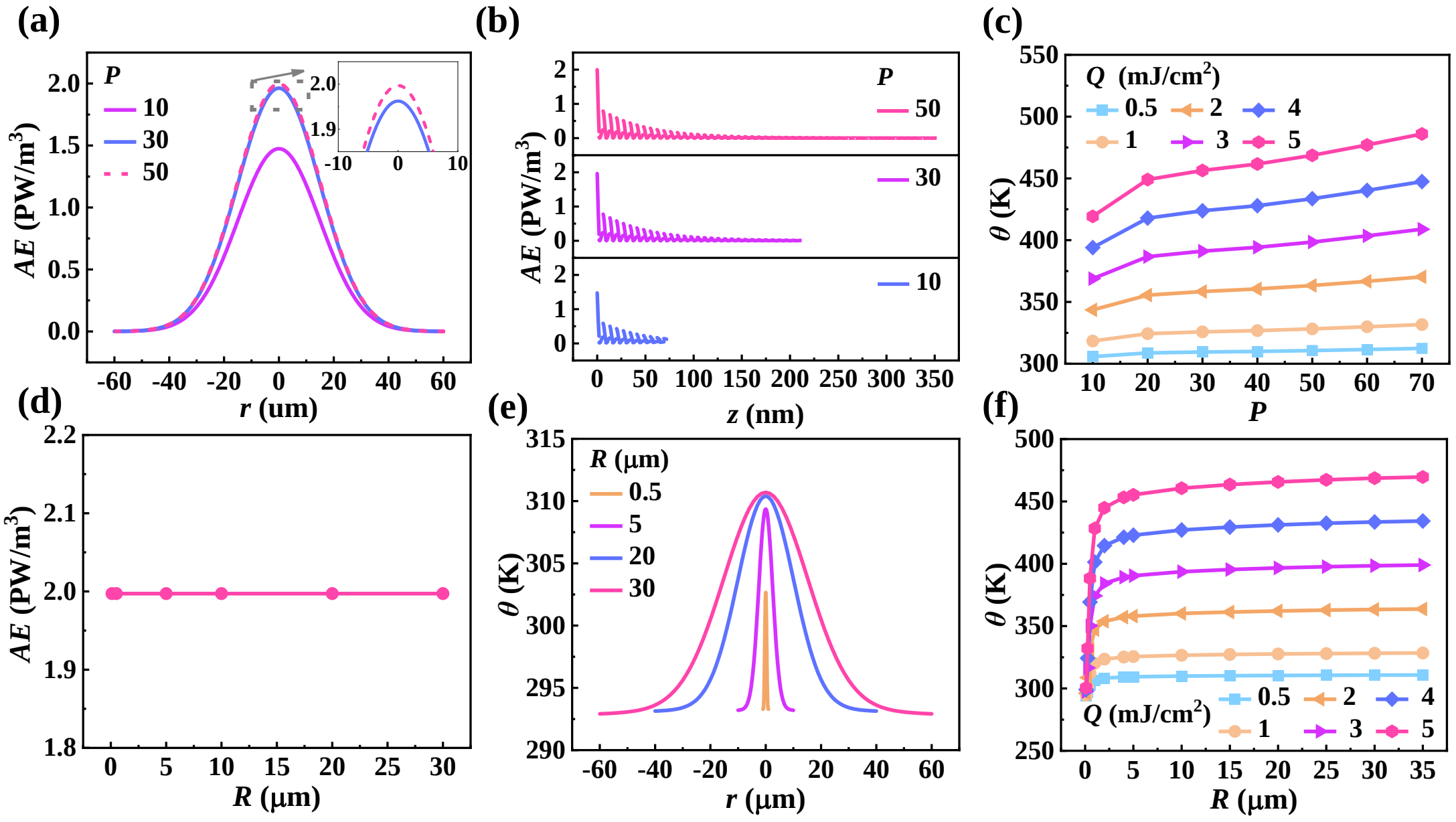

**Fig. 4 | Dependence of electromagnetic absorption and pulsed thermal response on period number and beam size.**

**a** Radial **electromagnetic** absorption profiles for multilayers with 10-50 periods under irradiation. All curves follow Gaussian shapes, and the profiles for 30 and 50 periods nearly overlap, indicating minimal radial variation beyond 30 periods. **b** Depth-dependent absorption distributions, with convergence emerging once the interference field becomes fully established for ≥30 periods. **c** Surface temperature under pulsed irradiation at different energy densities *Q*. Regardless of *Q*, the surface temperature increases monotonically with period number. **d** Radial absorption peaks for different beam radii *R*. **e** Radial temperature distributions under pulsed irradiation, with strong size dependence for small beams and weakened sensitivity for beams above a few micrometers. **f** Convergence of peak surface temperature with increasing beam radius.

is further quantified by the axial heat-flux evolution in Fig. 3(e). During the 8-ns pulse, the heat flux builds up globally, while in the cooling stage, the flux peak gradually shifts deeper into the multiple stacks. Critically, this observed peak shift is a direct manifestation of the coupled electromagnetic-thermal physics in the proposed ETCM, which is absent in conventional MHTM. The MHTM, by approximating the incident energy as a simple surface flux, can only predict a monotonically decaying heat-flux profile, which fails to capture the post-pulse migration of the flux peak. This observation highlights that thermal design of multilayer mirrors must account for the post-pulse redistribution of heat flux, as it can continue to drive heat into deeper layers even after the laser irradiation ceases. Significant heating can persist and even increase in deeper layers after the pulse, a delayed effect that the proposed ETCM uniquely captures. Therefore, the ETCM provides a more rigorous design framework for assessing thermal stability under pulsed EUV irradiation.

The preceding analysis validates the ETCM's capability to resolve the transient thermal response under pulsed EUV irradiation. For practical mirror design, it is crucial to identify the key structural and operational parameters that dictate the thermal load. As illustrated in Fig. 4(a), the surface absorption energy distribution first increases and then saturates with the period number ($P$) of multilayer. Specifically, a pronounced rise is observed when the period number increases from 10 to 30, beyond which the absorption gradually saturates. This saturation can be explained by examining the axial absorption distribution in Fig. 4(b). When the period number exceeds 30, the electromagnetic wave cannot penetrate to greater depths, resulting in negligible absorption in deeper layers. At this point, the optical interference field within the multilayer becomes fully developed. Consequently, adding further periods does not alter the internal electromagnetic field distribution, leading to a converged energy deposition profile. However, even though the absorption saturates beyond 30 periods, the heat accumulation continues to rise with the increasing period number. As shown in Fig. 4(c), the surface temperature under pulsed irradiation demonstrates a monotonic rise with increasing period number, regardless of the energy density. This rising trend indicates that a thicker multilayer poses a greater barrier to heat diffusion into the substrate, even though it exhibits a converged optical absorption profile and a stabilized reflectivity with increasing period number, as confirmed by Supplementary Fig. S5. The increased effective thermal resistance of the thicker multilayer impedes cooling, leading to stronger heat accumulation near the surface. Therefore, for optimal thermal management, the multilayer should be designed with the minimum period number necessary to achieve the converged optical interference field and thus the reflectivity, balancing optical performance with thermal dissipation.

In addition to structure periodicity, the beam size is another critical parameter governing the spatial distribution of the thermal load. As shown in Fig. 4(d), for a given incident energy density, the absorbed energy density at the surface center remains independent of the beam radius ($R$), confirming that optical absorption is solely governed by the local energy density, rather than by the absolute beam sizes. In contrast, the resulting surface temperature exhibits a pronounced size effect as illustrated in Fig. 4(e). For small beam radii, the peak surface temperature at the beam center increases rapidly with increasing $R$. However, this sensitivity diminishes markedly for larger beams, and the peak temperature saturates once the radius exceeds ~30 μm. This thermal size effect arises from the competition between in-plane radial heat spreading into surroundings and out-of-plane heat transfer into the Si substrate. Specifically, for small beam radii, the large surface-to-volume ratio enhances lateral heat spreading, allowing a significant fraction of the absorbed energy at the surface center to diffuse radially before being transferred into the substrate, thereby limiting local heat accumulation. As the beam diameter increases, the relative contribution of in-plane thermal diffusion gradually diminishes and out-of-plane heat transfer dominates, and heat deposited at the surface center becomes increasingly confined, with less opportunity to spread laterally prior to out-of-plane dissipation. Consequently, the surface temperature rises with beam size until a steady thermal confinement regime is reached, beyond which further increases in beam radius no longer affect the peak temperature. It is worth noting that the transition beam radius is dependent on the thickness of Mo/Si multilayers. Thus, beam radius is a key design parameter for the thermal stability and lifetime of Mo/Si mirrors.

Unlike the dependence on structure periodicity, the surface temperature peak always saturates with increasing beam radius across all energy densities varying from 0.5 to 5 mJ/cm$^2$, as shown in Fig. 4(f). This saturation indicates that once the beam radius exceeds a characteristic length scale, the surface-center temperature becomes insensitive to the lateral beam extent, as heat transport is dominated by out-of-plane conduction into the Si substrate rather than in-plane radial diffusion. Consequently, further increases in beam radius do not result in additional temperature rise at the beam center. Similar convergence behaviors observed under variations in multilayer period number, pulse repetition period, and pulse width are summarized in Supplementary Fig. S6. Additional insight into the underlying heat-transport mechanisms is provided by Supplementary Fig. S7, which shows that interfacial thermal resistance primarily affects interfacial temperature discontinuities, whereas higher energy density amplifies temperature gradients throughout the whole multilayer.

Although the present analysis is focused on Mo/Si multilayer mirrors under EUV

irradiation, the underlying electromagnetic–thermal coupling framework is not restricted to a specific material system or wavelength regime. The formulation relies only on (i) wave interference in stratified media, (ii) volumetric energy dissipation governed by electromagnetic attenuation, and (iii) transient heat conduction across layered structures. These ingredients are common to a broad class of multilayer nanostructures, including dielectric Bragg reflectors, X-ray and soft X-ray optics, multilayer metasurfaces, and optoelectronic stacks operating under pulsed laser excitation. By replacing the optical constants, layer geometry and excitation waveform, the same analytical framework can be directly extended to other material combinations and spectral ranges without reformulating the governing equations. The present Mo/Si–EUV system therefore serves as a representative platform that exposes general interference-controlled photothermal transport mechanisms in multilayer nanofilms, rather than a case-specific thermal model.

**Interfacial compaction-driven failure**

The preceding parametric study has identified the dependence of the temperature rise on key structural and operational parameters under EUV irradiation. These elevated temperatures under prolonged EUV irradiation can trigger various degradation mechanisms that threaten mirror performance and lifetime[40–43]. Among them, interfacial thermal compaction, which is driven by atomic interdiffusion and $MoSi_2$ formation at the interface as illustrated in Fig. 5(a), is identified as the most prevalent failure mode under EUV irradiation[42]. Therefore, we employ the validated ETCM to quantify this key damage process and its impact on optical reflectivity.

As shown in Supplementary Fig. S8, the reflectivity of the multilayer mirror decreases continuously as interfacial compaction progresses. This degradation occurs because the compaction reduces the individual layer thicknesses, disrupting the precise optical periodicity required to establish a stable interference condition for 13.5-nm EUV light. When the total compaction reaches approximately 12.8 nm, the reflectivity drops sharply to 26.6%, an exponential fraction of its initial peak value of ~72.3%. Thus, we define 12.8 nm (3.7% of the total mirror thickness) as the functional failure threshold for Mo/Si multilayers, below which the mirror cannot meet operational requirements. Using this defined threshold, the compaction model is integrated with ETCM to predict the evolution of interfacial compaction under prolonged EUV irradiation at different laser energy densities (see Supplementary Note 9 for details).

As shown in Fig. 5(b), the thickness ($w_f$) of $MoSi_2$ at the Mo/Si interface increase stepwise over five consecutive pulse cycles, directly following the transient temperature evolution induced by pulsed EUV irradiation. Each temperature spike drives a rapid increase in $MoSi_2$ formation, while a much slower growth occurs during the thermal relaxation between pulses. Figure 5(c) further illustrates the long-term evolution of $w_f$ and the interdiffusion parameters governing the $MoSi_2$ growth, including the effective activation energy $E_A$ and the diffusion coefficient $D_0$ for (Supplementary Note 9), over a 10-year timescale. Since both $E_A$ and $D_0$ depend on $w_f$, and $w_f$ is the integral of the growth rate determined by $E_A$ and $D_0$ (Eqs. S(3)-S(5)), $w_f$ exhibits a rapid initial increase followed by slower growth, whereas $E_A$ gradually rises over time. The early-stage acceleration of $w_f$ is attributed to the initially low $E_A$, which progressively increases as interfacial $MoSi_2$ accumulates, thereby reducing the subsequent growth rate of $w_f$ (Supplementary Fig. S9). Notably, as illustrated in Supplementary Figs. S10 and S11, compaction values calculated using the peak temperature of pulse cycle systematically overestimate the actual $w_f$. In contrast, calculations based on the cycle-averaged temperature provide a more accurate representation of the effective thermal driving force for interdiffusion. Therefore, in all long-term lifetime predictions presented in this work, the cycle-averaged temperature is employed to evaluate interfacial compaction under prolonged EUV irradiation.

Figure 5(d) shows the axial and radial compaction distribution after 10-year irradiation at 5 mJ/cm$^2$. Clearly, compaction mainly occurs within the upper several

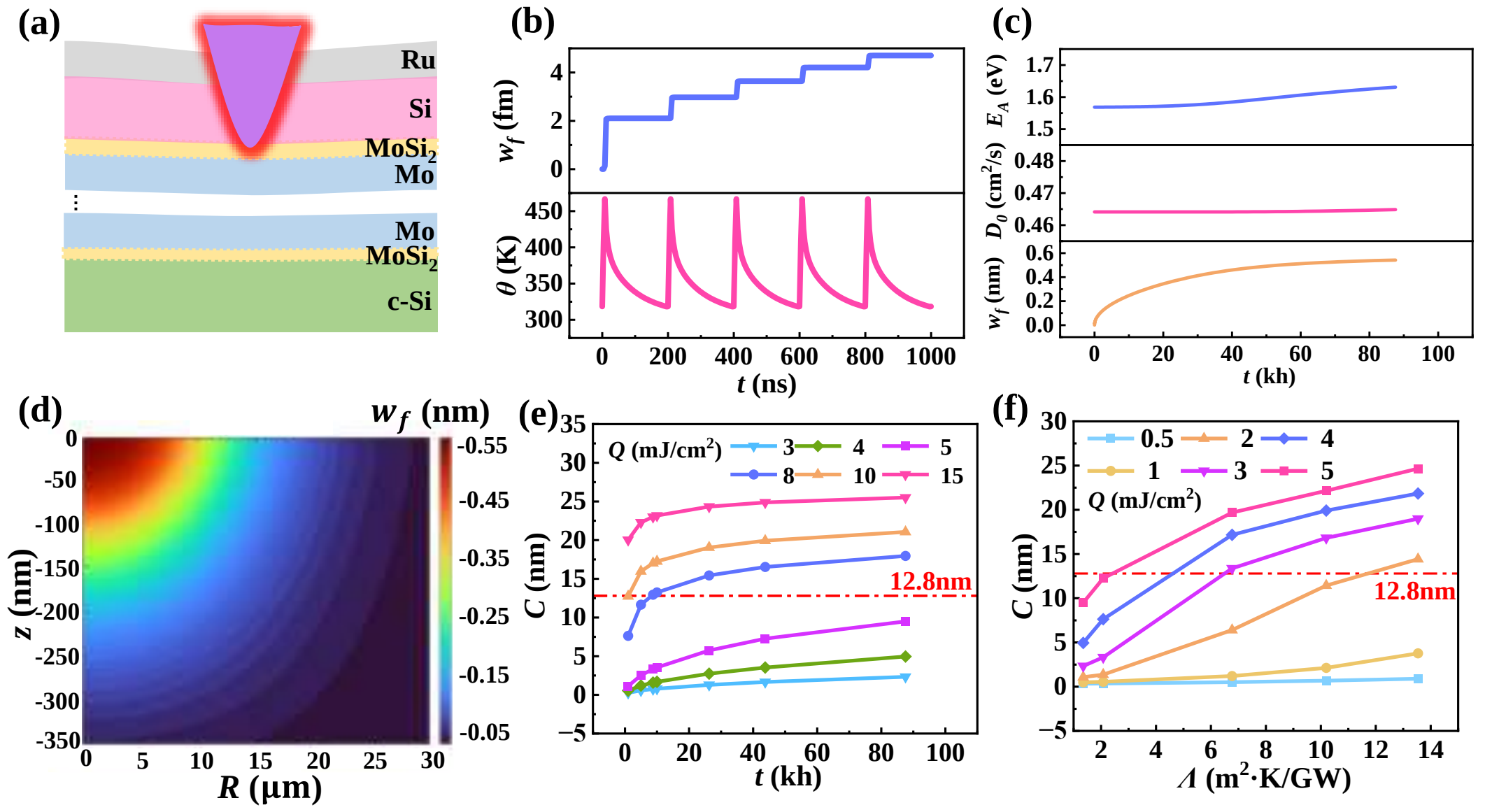

**Figure 5 | Interfacial compaction-driven failure mechanism of Mo/Si multilayers under EUV irradiation.**

**a** Schematic illustration of interfacial $MoSi_2$ formation under EUV irradiation. **b** Temporal evolution of $MoSi_2$ formation and temperature over the first five pulse cycles (initial activation energy $E_A$=1.58 eV, interfacial thermal resistance $\Lambda$=1.35 m$^2$·K/GW, energy densitiy $Q$=5 mJ/cm$^2$). **c** Temporal evolution of $MoSi_2$ formation and the associated interdiffusion parameters $D_0$ and interdiffusion activation energy $E_A$ over 10 years. **d** Spatial distribution of $MoSi_2$ formation. **e** Evolution of interfacial compaction $C$ under different irradiation density $Q$. **f** Effect of interfacial thermal resistance on interfacial compaction.

periods of the multilayer mirror and decays significantly along the axial direction. This non-uniform distribution directly reflects the distribution of electromagnetic field as shown in Fig. 3(a), which deposits most of its energy within the upper several periods of the multilayer. This concentrated energy absorption causes a pronounced temperature rise specifically at these upper interfaces, leading to the accelerated atomic interdiffusion and $MoSi_2$ formation at these locations.

During EUV irradiation, the energy density can vary substantially. Fig. 5(e) compares the time dependent compaction behaviors at different energy densities. At low energy densities below 5 mJ/cm$^2$, the total compaction remains below the failure threshold (12.8 nm) even after 10 years (87.6 kh) under continuous irradiation. In contrast, when the energy density exceeds 8 mJ/cm$^2$, the failure threshold is reached within 10 kh. In particular, for energy densities above 10 mJ/cm$^2$, failure occurs almost immediately during the initial stage of irradiation.

Since the thickness of interfacial $MoSi_2$ continuously evolves during irradiation, the interfacial thermal resistance should be treated as a variable parameter rather than a fixed constant. To systematically evaluate its effect, we investigate how different interfacial thermal resistances influence temperature evolution and interfacial compaction. The interfacial thermal resistance used here is obtained from our molecular dynamics-based calculations[44] and is consistent with reported literature values[45]. Figure 5(f) shows the predicted interfacial compaction at 10 years for different interfacial thermal resistances ($\Lambda$). The results indicate that a higher interfacial thermal resistance leads to increased heat accumulation and higher local temperatures, which in turn accelerates atomic interdiffusion and promotes interfacial compaction. Consequently, the lifetime of Mo/Si mirrors and the energy density threshold of laser irradiation decrease monotonically with increasing $\Lambda$.

## Conclusion

In conclusion, we have established a unified analytical framework that self-consistently couples electromagnetic interference with transient heat transport in Mo/Si multilayer mirrors under pulsed EUV irradiation. By explicitly resolving interference-governed volumetric absorption, the model reveals thermal behaviors that are fundamentally inaccessible to conventional surface-heating descriptions, including pronounced axial temperature gradients and a delayed post-pulse migration of the heat-flux maximum into deeper layers. This interference-driven photothermal response leads to a fundamental trade-off in multilayer design: while increasing the period number is essential for achieving high reflectivity, it simultaneously enhances temperature by impeding heat dissipation. The model further identifies scaling relations linking multilayer thickness, beam size, and interfacial thermal resistance to heat accumulation, providing quantitative guidance for thermal–optical optimization. Importantly, the framework rigorously reduces to the traditional multilayer heat conduction model in the surface-absorption limit, establishing a unified theoretical description across modeling regimes. By incorporating interfacial compaction kinetics, we further connect transient thermal response to long-term structural degradation and lifetime limits. Beyond EUV optics, the concepts and methodology introduced here offer a general route for understanding and designing interference-controlled photothermal transport in multilayer nanostructures exposed to intense pulsed radiation.

**Data availability**

The raw data that support the findings of this study are available from the corresponding author upon reasonable request.

**Code availability**

The codes used in this paper are available from the corresponding authors upon request.

**Acknowledgments**

This work was supported by National Natural Science Foundation of China (Grant no. 52206092), the National Key Research & Development Program of China (Grant No. 2024YFF0508900), Basic Research Program of Jiangsu (BK20250035), and the Big Data Computing Center of Southeast University and the Center for Fundamental and Interdisciplinary Sciences of Southeast University.

**Contributions**

C.H.L. conceptualized the work. C.H.L and Y.Y. supervised the work. H.Y.H and L.M. did the calculation. H.Y.H and C.W. performed the theoretical analysis. H.Y.H., C.W., and L.M. wrote the manuscript. All authors discussed results and edited the text.

**Declaration of interests**

The authors declare that they have no competing interests.